\documentclass[preprint,amsmath,amssymb,aps,prl,longbibliography,lengthcheck]{revtex4-1}
\usepackage{graphicx}
\usepackage{dcolumn}
\usepackage{bm}
\usepackage{hyperref}
\usepackage{color}
\usepackage{soul}
\begin{document}

\preprint{PRC/Preprint 5.4}

\title{Measurements of the $^{48}$Ca($\gamma,n$) Reaction}

\author{J.~R.~Tompkins}
\author{C.~W.~Arnold}
\author{H.~J.~Karwowski}
\author{G.~C.~Rich}
\affiliation{Department of Physics and Astronomy, University of North Carolina, Chapel Hill, NC 27599}
\affiliation{Triangle Universities Nuclear Laboratory, Durham, NC 27708}

\author{L.~G.~Sobotka}
\affiliation{Departments of Chemistry and Physics, Washington University, St. Louis MO, 63130}

\author{C.~R.~Howell}
\affiliation{Department of Physics, Duke University, Durham, NC 27708}
\affiliation{ Triangle Universities Nuclear Laboratory, Durham, NC 27708}

\date{\today}

\begin{abstract}
The $^{48}$Ca($\gamma$,$n$) cross section was measured using $\gamma$-ray beams of energies between 9.5 and 15.3 MeV generated at the Triangle Universities Nuclear Laboratory (TUNL) high-intensity $\gamma$-ray source (HI$\gamma$S). Prior to this experiment, no direct measurements had been made with $\gamma$-ray beams of sufficiently low energy spread to observe structure in this energy range. The cross sections were measured at thirty-four different $\gamma$-ray energies with an enriched $^{48}$Ca target. Neutron emission is the dominant decay mechanism in the measured energy range that spans from threshold, across the previously identified M1 strength, and up the low-energy edge of the E1 giant dipole resonance (GDR). 
This work found $B(M1)$ = 6.8 $\pm$ 0.5 $\mu_{N}^2$ for the 10.23 MeV resonance, a value greater than previously measured.
Structures in the cross section commensurate with extended random-phase approximation (ERPA) calculations have also been observed 
 whose magnitudes are in agreement with existing data.
\end{abstract}

\maketitle

\section{Introduction}
The investigation of magnetic dipole transitions provides information about the spin and isospin parts of the effective nuclear interaction. It has been observed that the strength distribution of these M1 transitions is fragmented in nuclei. Due to recent computational progress, shell-model studies performed in large model spaces are able to describe details of the fragmentation of this mode in $pf$-shell nuclei. $^{48}$Ca is an excellent test case for these studies because of the relatively small fragmentation of the strength distribution and its simple shell structure. 

In the independent-particle model, an M1 transition in $^{48}$Ca is a pure neutron spin-flip excitation from $0f_{7/2} \rightarrow 0f_{5/2}$. In this extreme model, the transition strength is proportional to the number of unmatched spin-orbit nucleons and equal to $\vert\langle j$=$\frac{7}{2} \|$M1$\|j$=$\frac{5}{2}\rangle\vert^{2}=$ 11.98 $\mu _{N}^{2}$ \cite{Bohr-Mottelson}. It has been known for 25 years that the strength distribution of the M1 transition, which is fragmented throughout the low-energy region, is dominated by a single fragment at 10.23 MeV. This fragment has been observed in ($e$,$e^{\prime}$) \cite{Steffen83}, ($p$,$p^{\prime}$) \cite{Berg82,Rehm82,Fujita82,Crawley83,Tamii071}, and ($p$,$n$) as an isobaric analog in $^{48}$Sc \cite{Anderson80}. The interest in this excitation is not in the location of the major fragment but in its strength \cite{Richter85,Richter00}, measured by ($e$,$e^{\prime}$) to be B(M1)\ = 3.9$\pm $0.3 $\mu _{N}^{2}$ \cite{Steffen83}. This strength amounts to approximately 1/3 of that expected in the independent-particle model. In addition, the summed strength $\Sigma$ B(M1) in the region between 7.7 and 12.7 MeV has been measured to be 5.3$\pm $0.6 $\mu _{N}^{2}$, which accounts for less than 1/2 of the predicted strength. The general fragmentation, localization at 10.23 MeV, and overprediction of the $\Sigma$ B(M1) by an independent particle model in the low-energy region of the M1 strength distribution all indicate the extent and nature of the complex correlations that exist in nuclei. 

Quenching of the M1 strength is not surprising as standard intra-shell configuration mixing (CM) will cause destructive interference between the $0f_{7/2}\rightarrow 0f_{5/2}$ and $0f_{5/2} \rightarrow 0f_{7/2}$
contributions to the 1p1h excitation \cite{Arima54}. This quenching at the 0$\hbar \omega$ level amounts to no more than 25\% and is thus less than half of that required to explain the experimental results
cited above. Additional quenching has been found in studies of higher-order CM extending beyond
the $pf$-shell \cite{Takayanagi88,Brand88,Brand90,Brand90t,Kamerdzhiev1989,Kamerdzhiev1993}. These works have indicated that mixing at the 2$\hbar \omega$ level is significant, and while contributions at the 4$\hbar \omega $ level are small, they are not entirely negligible \cite{Takayanagi88}. Altogether, the inclusion of these effects improves, but fails to bring, agreement between theory and experiment. Consequently, other sources of quenching and the relocation of M1 strength have been proposed \cite{Richter85, Takayanagi88}, these include: CM at the 6$\hbar \omega $ level, the effects of short-range correlations, $\Delta (1232)$-nucleon hole coupling, and $\rho$-meson exchange. All effects beyond the 0$\hbar \omega$ level are dealt with phenomologically in shell model calculations by rescaling the free spin operator g$_{s}^{free}$.

For the E1 strength, RPA calculations in the region of the low-energy edge of the GDR show multiple sharp peaks in the E1 strength distribution \cite{Brand90}. 
These peaks, which are artificially smoothed for presentation in standard RPA, broaden naturally in extended random-phase approximation (ERPA) calculations due to the inclusion of coupling to 2p2h intra- and intershell configurations: the more coupling, the smoother the response. 
Green's function methods extending the theory of finite Fermi systems (TFFS) \cite{Migdal67} also predict this smoothing when coupling of 1p1h states to more complex configurations is included \cite{Kamerdzhiev1989}. 
Thus the presence and magnitude of the structures in the E1 strength distribution reflect the importance of such 2p2h configurations to the ground-state configuration.

The present work uses real photons to excite M1 and E1 transitions in an enriched $^{48}$Ca sample. The ($\gamma ,n$) cross section was measured from $E_{\gamma}$ = 9.5 MeV, just below the neutron emission threshold, to 15.3 MeV using a highly-efficient Model IV Inventory Sample Counter (INVS) \cite{Arnold09a} that was adapted for beam experiments. The energy resolution of the beam, which was better than 2.6\%, is far superior to that achievable in previous photon work that utilized a bremsstrahlung spectrum  \cite{OKeefe87} and provides the ability to resolve the structures predicted by RPA calculations. At these energies, single-neutron emission is the dominant decay channel and thus deduction of a B(M1) for the strong M1 fragment at 10.23 MeV as well as inferences about the E1 strength distribution can be made.

\section{Experimental Setup}

\begin{figure*}[tbp]
\begin{center}
\includegraphics[scale=0.6]{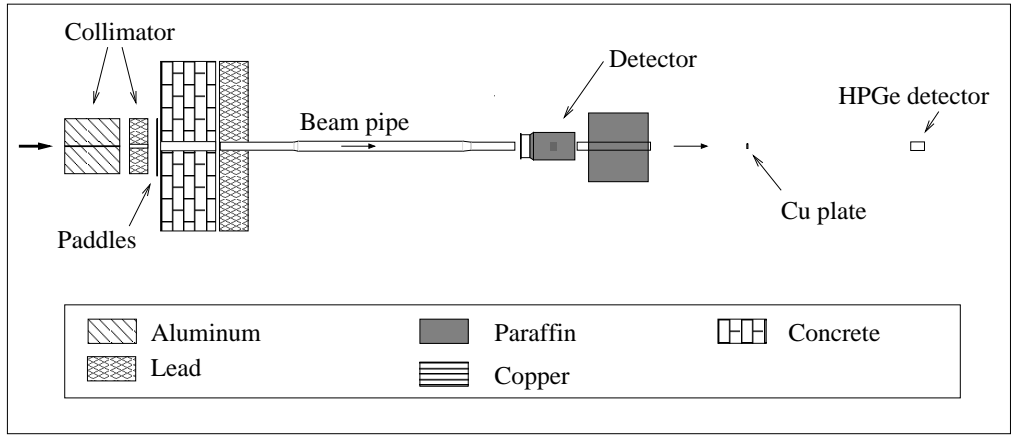}
\end{center}
\caption{Scaled diagram of the $^{48}$Ca$(\protect\gamma,n)$ experimental setup. The $\gamma$-ray beam direction is indicated by the arrows.}
\label{expsetup}
\end{figure*}

\subsection{The High Intensity $\gamma$-ray Source}

The high-intensity $\gamma$-ray source (HI$\gamma $S) at Triangle Universities Nuclear Laboratory (TUNL) produced nearly monoenergetic $\gamma$-ray beams by means of Compton backscattering inside the optical cavity of a storage ring based free-electron laser. Details pertaining to the general operation and capabilities of the facility can be found in Ref. \cite{Weller09}. The beams were circularly polarized and produced with both high, $\Delta E_{\gamma}/E_{\gamma}=0.9-1.6$\%, and low, $\Delta E_{\gamma}/E_{\gamma}=1.9-2.3 $\%, energy resolutions. The low resolution beams were used to measure the $^{48}$Ca$(\gamma ,n)$ cross section in the energy region $E_{\gamma}=9.5-15.3$ MeV in 0.25 MeV steps. Energies at which structure in the cross section was observed were remeasured using high resolution beams and 0.1 MeV steps.

The $\gamma$-ray beams were collimated 60 m downstream from the electron storage ring by a circular aperature of diameter 0.95 cm in a 60 cm long block of aluminum. Aluminum was chosen to be the material for the primary collimator because it has a high neutron separation energy, $S_{n}$ = 13 MeV. A lead clean-up collimator with a 2.81 cm aperature was positioned downstream of the Al collimator to decrease the flux of small-angle scattered $\gamma$ rays from reaching the target, see Fig \ref{expsetup}.

\subsection{The INVS and Targets}

\begin{figure}[b]
\includegraphics[scale=0.9]{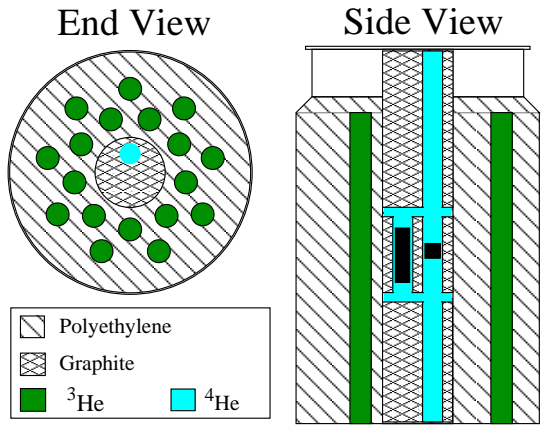}
\caption{(Color online) End and side cut-away views of the INVS showing the $^3$He tubes arranged in two concentric rings around a central cavity filled with 3 graphite cylinders. The central cylinder holds the four targets (black rectangles) and rotates about the symmetry axis of the detector to change targets.}
\label{ndetector}
\end{figure}

Four targets ($^{48}$Ca, $^{nat}$Ca, D$_{2}$O, and H$_{2}$O) were located in a revolver-like holder in the central cavity of the Model IV INVS counter. The holder was positioned near the center of the INVS to maximize the neutron detection efficiency, which has been measured to be as high as 60\% for neutrons with energies below 1 MeV \cite{Arnold09a}. The detector consists of 18 $^{3}$He proportional-counting tubes embedded in a polyethylene moderator, as shown in Fig. \ref{ndetector}. All of the tubes are filled with 6 atm of $^{3}$He gas. The tubes are arranged into two concentric rings with 9 tubes in each. The radius of the inner ring is 7.2 cm and that of the outer ring is 10.6 cm. Three TTL outputs are produced by the INVS, one for each of the rings and one for the logical OR of the two.

Measurements of $^{48}$Ca($\gamma$,$n$) were obtained with a 92.4\% enriched, 2.72 g $^{48}$Ca target. A 2.54 cm diameter plexiglass capsule filled with argon protected the target against oxidation and low-density foam centered the 1.27 cm diameter x 1.19 cm long calcium target within the capsule. Contributions to the neutron yield from target impurities and the casing were measured using a 2.29 g $^{nat}$Ca target in an identical casing. The entire data set employs $\gamma$ rays below the neutron emission threshold of $^{40}$Ca, $S_n = 15.6$ MeV.

A 99.9\% pure D$_{2}$O target, encased in a 7.5 cm long, 1.92 cm outer diameter polyethylene cylinder with 25.4 $\mu$m thick walls, was used for measurements of the $\gamma$-ray flux by means of the $^2$H($\gamma,n$) reaction. Background contributions from $^1$H($\gamma,\gamma^{\prime}$) to this measurement were measured at each $\gamma$-ray beam energy using a second target of identical geometry filled with deionized H$_{2}$O.


\subsection{Beam Diagnostics}
The energies of the $\gamma$-ray beams were measured at 0.5 MeV intervals using a high-purity germanium detector positioned in the beam. For these measurements, the beams were attenuated to less than 4 $\times$ $10^{3}$ $\gamma /s$ by precision-machined copper blocks prior to collimation. The detector was calibrated with a $^{60}$Co source and confidence in the extrapolated calibration was established by measurement of the sharp M1 resonance at 10.23 MeV.

Relative normalization of the flux between different target measurements, which were not concurrent, was made possible by plastic scintillating paddles that remained in the beam at all times. 


\section{Analysis}

\subsection{B($\mathbf{M1;0 \rightarrow 10.23}$ MeV)}
The B(M1) strength can be extracted from the integrated ($\gamma ,n$) cross section across the resonance by assuming that contributions to the cross section are dominated by the M1 resonance. Because the energy width of the $\gamma$-ray beam is roughly an order-of-magnitude larger than the upper limit of the resonance width, 17 keV \cite{Tamii071}, transitions nearby are unavoidably excited. A second requirement is that the flux per unit energy across the width of the resonance is constant. This condition is satisfied because the centroid of the beam for one data set was at the resonance energy and the resonance is sufficiently narrow compared to the energy width of the beam.
Finally, the decay mode must be dominantly neutron emission, which is generally true when a nucleus is excited above the particle separation energy. The primary source of uncertainty is in the differential flux of the $\gamma$-ray beam at the resonance energy.

The B(M1) can be derived from the rate of neutrons detected from the $^{48}$Ca($\gamma,n$) reaction, $R$, where
\begin{eqnarray}
	R = n_{48} \int_{\Delta_{res}} \sigma \epsilon \frac{\mathrm{d}\phi }{\mathrm{d}E_{\gamma}}  \mathrm{d}E_{\gamma}.
\end{eqnarray}
The quantity $\epsilon$ is the efficiency for detecting neutrons, $n_{48}$ is the areal density of $^{48}$Ca nuclei exposed to the $\gamma$-ray beam, $\sigma$ is the reaction cross section, and $\mathrm{d}\phi /\mathrm{d}E_{\gamma}$ is the differential flux of photons with respect to $\gamma$-ray energy. The integration is over the resonance width, $\Delta_{res}$. Using both of the assumptions, the equation can be put into the form
\begin{equation}
	\frac{R}{n_{48} \epsilon \left(\mathrm{d}\phi/\mathrm{d}E_{\gamma}\right) |_{E_{res}}} = \int_{\Delta_{res}} \sigma \mathrm{d}E_{\gamma}.
	\label{a}
\end{equation} 
As discussed in \cite{Bohr-Mottelson}, the B(M1) is related to the right hand side of Eq. (\ref{a}) by 
\begin{equation}
	\int_{\Delta_{res}} \sigma\, \mathrm{d}E_{\gamma} = 4.41 \times 10^{-3}\, E_{res} B(M1)\quad \mathrm{ fm^2\, MeV}.
\label{b}
\end{equation}
Substitution of Eq. (\ref{b}) into Eq. (\ref{a}), yields 
\begin{equation}
	B(\mathrm{M}1) = \frac{(2.27 \times 10^2 \mathrm{\mu_{N}^{2}/fm^{2}}) \, R}{E_{res} n_{48} \left( \mathrm{d}\phi /\mathrm{d}E_{\gamma} \right)|_{E_{res}} \epsilon}. \label{BM1equation}
\end{equation}
The differential flux at the resonance energy, $\mathrm{d}\phi / \mathrm{d}E_{\gamma} |_{E_{res}}$, is determined from the measured $\gamma$-ray beam energy profile and total flux by
\begin{equation}
	\frac{\mathrm{d} \phi}{ \mathrm{d}E_{\gamma}} \bigg|_{E_{res}} = \frac{\mathrm{d}\phi^{p}}{\mathrm{d}E_{\gamma}} \bigg|_{E_{res}} \frac{\phi_{tot}^d}{\phi_{tot}^{p}}.
\end{equation}
The quantity $\mathrm{d}\phi^{p} /\mathrm{d}E_{\gamma}$ is the measured beam energy profile after corrections for an energy dependent efficiency and detector response have been applied. It is scaled by the ratio of the total flux as measured with the $^2$H($\gamma,n$) reaction, $\phi_{tot}^{d}$, to the integrated energy profile, $\phi_{tot}^{p}$. The value of each parameter in the B(M1) calculation along with the result is presented in Table \ref{bm1result}.

\begin{table}
\caption {\label{bm1result}
The data used in the calculation of the B(M1;$\, \rightarrow \,$10.23 MeV) along with the result are included in the table.}
\begin{ruledtabular}
\begin{tabular}{lcr}
\textrm{Quantity}& \textrm{Value} & \textrm{Units} \\ 
\colrule
$n_{48}$ & 1.92 $\times$ 10$^{22}$ & cm$^{-2}$ \\  
$\epsilon$(0.28 MeV) & 0.58 $\pm$ 0.03 & abs \\ 
R & 610 $\pm$ 20 & cts/s \\ 
$\mathrm{d}\phi / \mathrm{d}E_{\gamma}\Big|_{E_{\gamma}=E_{res}}$ & (3.49 $\pm$ 0.12) $\times$ 10%
$^4$ & $\gamma/s/keV$ \\ 
B(M1;$0 \rightarrow \mathrm{10.23 MeV}$) & 6.8 $\pm$ 0.5 & $\mu_N^2$ \\ 
\end{tabular}
\end{ruledtabular}
\end{table}

\subsection{Cross Section Determinations}

\begin{figure}[b]
\includegraphics[scale=0.47]{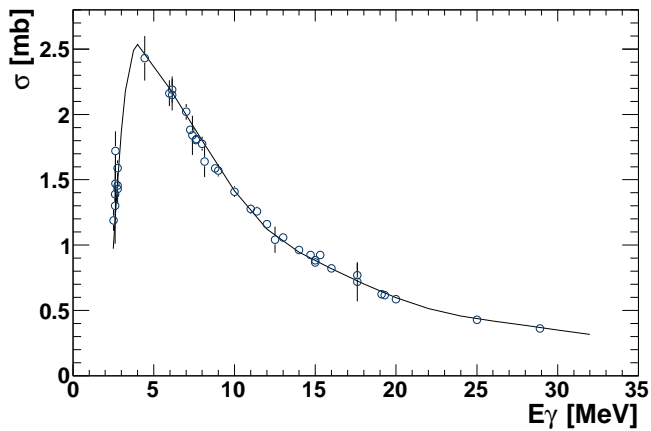}
\caption{(Color online) Theoretical calculations of the $^2$H$(\gamma,n)$ total
cross section (solid curve) and data (open circles) (figure taken from \cite{Schiavilla05}). The solid curve represents the overlay of the results of five realistic calculations, which are shown indistinguishable.}
\label{dgncs}
\end{figure}

The $^2$H$(\gamma,n)$ reaction was used to measure the absolute flux of the incident $\gamma$-ray beam. Calculations of the total cross section using realistic nucleon-nucleon potentials, such as CD-Bonn, Nijmegen-1, AV18, AV8, and AV6, are considered to be well understood and reliable. For example, the CD-Bonn potential fits world pp(np) data below 350 MeV, up until the year 2000, to a $\chi^2$ per datum of 1.01(1.02) \cite{Machleidt01}. Further confidence is gained from the fact that all the potentials produce indistinguishable values for the $^2$H($\gamma,n$) cross section \cite{Schiavilla05} in the energy region of interest, see Fig. \ref{dgncs}. For these reasons, a polynomial fit to the CD-Bonn cross sections was used in this analysis.

\begin{figure}[tbp]
\begin{center}
\includegraphics[scale=0.43]{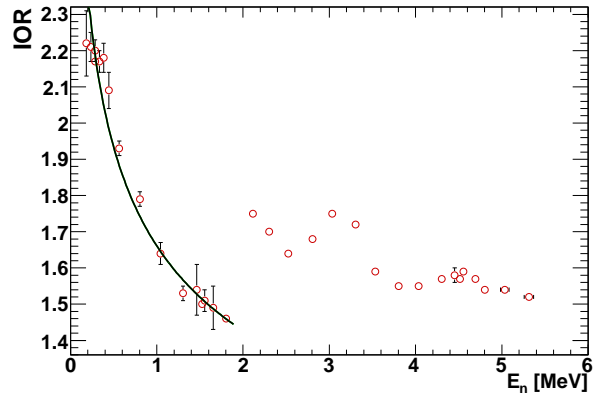}
\end{center}
\caption{(Color online) An energy dependent fit (solid curve) to the $^{48}$Ca IOR (open circles) from $S_n$ to 12 MeV, the energy at which decay to the first excited state in $^{47}$Ca becomes energetically allowed, was used to determine average neutron energies.}
\label{fittedior}
\end{figure}

The $^{48}$Ca($\gamma$,n) cross section is determined relative to the $^2$H$(\gamma,n)$ cross section as follows

\begin{center}
\begin{eqnarray}
\sigma_{48} = \frac{N_{48}} {N_{d}} \frac{P_{d}} {P_{48}} 
\frac{\epsilon_{d}} {\epsilon_{48}} \frac{n_{d}} {n_{48}} \sigma_{d}
\label{c}
\end{eqnarray}
\end{center}

\noindent where $\sigma$ is the total photodisintegration cross section, $N$ and $P$ are the total counts in the INVS and the paddle, respectively, corrected for background counts, $\epsilon$ is the neutron detection efficiency, and $n$ is the areal density of the target exposed to the beam. The subscripts $d$ and $48$ correspond to the $^2$H and $^{48}$Ca target nuclei, respectively. No correction to these data for the energy shape of the $\gamma$-ray beam has been applied and thus each point is to be considered an average cross section over the width of the beam.

The INVS was characterized with both simulation and experiment \cite{Arnold09a}. The neutron detection efficiencies were measured experimentally for energies below 2 MeV and used to validate the simulations done with the \textsc{MCNPX} code \cite{Hendricks05}. The agreement between these enabled the confident extrapolation of the simulated efficiencies for low energy neutrons to those with higher energy, such as the neutrons produced by the $^2$H($\gamma,n$) reaction. The characterization procedures were tailored to account for differences in the neutron angular distributions and energetics between the $^{48}$Ca($\gamma,n$) and $^2$H($\gamma,n$) reactions. 

The emitted neutrons from the photodisintegration of $^{48}$Ca at $E_{\gamma} > 12.0$ MeV are not monoenergetic. However, the energy distribution is well represented by the average $\langle E_n \rangle$ and a detection efficiency corresponding to $\langle E_n \rangle$ was used. This efficiency was determined from the \textsc{MCNPX} simulation and defined as $\epsilon = n_{dis}/n_{ini}$ where $n_{dis}$ is the number of neutron disappearances in the active volume of the $^{3}$He tubes and $n_{ini}$ is the number of initial neutrons. The ratio of neutron counts in the inner ring to those in the outer, now referred to as the inner-to-outer ratio (IOR), was used to determine $\langle E_n \rangle$. For $E_{\gamma} < 12.0$ MeV, the energy dependence of the ratio could be fit with a single power law, see Fig. \ref{fittedior}, whose inversion was used to determine $\langle E_n \rangle$ for all relevant $E_{\gamma}$.

\begin{center}
\begin{figure}[tbp]
\includegraphics[scale=0.40]{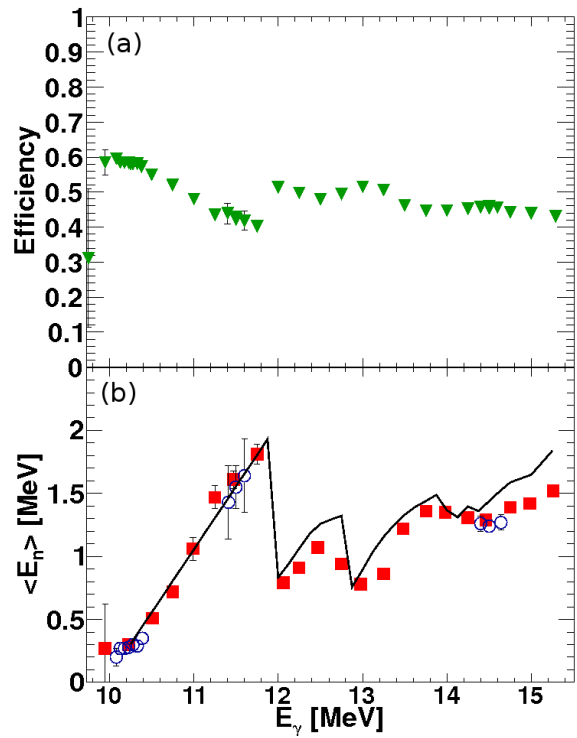}
\caption{(Color online) (a) The efficiency for detecting neutrons from the $^{48}$Ca($\gamma,n$) reaction at a given $E_{\gamma}$ is plotted (triangles). (b) The $\langle E_n \rangle$ at each $\gamma$-ray beam energy as determined by \textsc{GEMINI++} (solid curve) and using the IOR technique for the high (circles) and low (squares) resolution data.  Uncertainties presented reflect statistical uncertainties only.}
\label{iorresult}
\end{figure}
\end{center}

The $\langle E_n \rangle$ determined with the IOR technique described above was validated using the statistical model code \textsc{GEMINI++} \cite{Charity08}. In this code, the decay of an excited nucleus with $J ^{\pi} = 1^-$ to all energetically allowed, known states with $J \leq 9/2$ in $^{47}$Ca \cite{Burrows06} was calculated using the Hauser-Feshbach formalism.  Decays of excited 1$^+$ states were not considered due to their very small M1 transition strengths \cite{Steffen83}. The fractional contribution of each decay channel was computed and used with the corresponding detection efficiencies to form the weighted average of the neutron detection efficiency at a given $E_{\gamma}$. The result was compared with that of the IOR technique and reasonable agreement was found, see Fig. \ref{iorresult}. Further validation using the full reaction model code, \textsc{TALYS} \cite{Talys}, produced similar results.
A relative uncertainty of $\pm$5 \% is assigned to the efficiency determination due to the assumption about the angular distribution and the fact that the neutron detector efficiency is angular dependent when $E_n > 0.5$ MeV \cite{Arnold09a}.

The procedure for characterizing the detector's response to neutrons from deuteron photodisintegration accounted for the known $\sin ^2 \theta$ angular distribution \cite{Stephenson87}. The associated kinematics were accounted for with the following equation,
\begin{equation}
		\epsilon(E_{\gamma}) = \int_0^{\pi} \epsilon \left[\theta, E_n(E_{\gamma}, \theta)\right] W(\theta) d\theta 
\end{equation}
where $W(\theta) = \sin ^2 \theta$, $\theta$ is the polar angle measured with respect to the $\gamma$-ray beam, and $E_n(E_{\gamma}, \theta)$ is the energy of the emitted neutron. The angle and energy dependent efficiencies were the result of simulated monoenergetic, conical sources of neutrons for angles $0 < \theta < \pi$ for all neutron energies. A relative uncertainty of $\pm$3 \% is assigned to the neutron detection efficiency as it pertains to this reaction in concordance with Ref.~\cite{Arnold09a}.

\begin{figure}[t]
\includegraphics[scale=0.45]{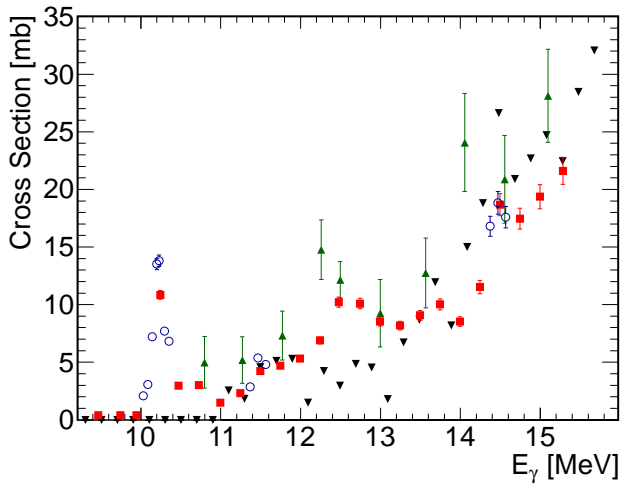}
\caption{(Color online) $^{48}$Ca($\protect\gamma$,n) total cross section measured with low (squares) and high (circles) resolution $\gamma$-ray beams. The uncertainties reflect the contributions of systematic uncertainties in the neutron detection efficiency. The upward pointing triangles are the data of Ref.~\cite{OKeefe87} and the inverted triangles are the converted data of Ref.~\cite{Strauch99}.}
\label{xsplot}
\end{figure}

The $^{48}$Ca($\gamma,n$) cross section determined using Eq. (\ref{c}) as described above is shown in Fig. \ref{xsplot}. The errors on the data points include statistical uncertainties as well as systematic uncertainties in the efficiency for detecting neutrons from the $^{48}$Ca($\gamma,n$) reaction.  

\section{Results and Discussion}

This work presents a new technique to extract localized M1 strength. As M1 and E1 are indistinguishable by the detection method, prior knowledge of the distribution of strength in the region and its magnitude is required. Any sizeable contribution to neutron counts due to strength near the major M1 resonance or within 9 keV of the peak, thus remaining unresolved by any measurement to date, weakens the validity of the B(M1) value extracted by this technique. However, the case of $^{48}$Ca is ideal because high-resolution ($p$,$p^{\prime}$) experiments show excitable transitions within the experimental energy spread at E$_{\gamma}$ = 10.23 MeV to be small. 
The relative contributions of these transitions on either side of the resonance, see Fig. 12 of Ref.~\cite{Tamii09}, have been estimated using a simple model of the spectrum using 3 gaussians, and their combined contribution was found to be on the order of 1-2\%, see Table \ref{GrassContributionTable}. The validity of the extraction method is thus retained.

The applicability of this technique is also dependent on $\Gamma_{\gamma} << \Gamma_n$ because this assumption is implied in Eq.~(\ref{b}). In $^{48}$Ca, the radiative decay width of the excited state has a Moszkowski estimate \cite{Moszkowski65} of 35 eV when considering deexcitation from the $0f_{5/2}$ to the $0f_{7/2}$ level. 
It is clear that even if the true width of the peak were an order of magnitude narrower than the 17 keV upper limit, the radiative partial width would be smaller than the total width by a factor of 50. 

\begin{table}
\caption{\label{GrassContributionTable} 
Estimation of peak contributions to the total neutron counts at the base of the large M1 resonance. This estimation includes weighting of the peak heights taken from Fig. 12 of Ref.\cite{Tamii09} by the intensity profile of the $\gamma$-ray beam. These estimates are upper limits because it has been assumed that any excitation of these levels results unambiguously in a neutron emission. }
\begin{ruledtabular}
\begin{tabular}{cc}
\textrm{$\mathbf{E_{\gamma}}$ (MeV)} 	&\textrm{Fractional Contribution} \\
\colrule
10.15 & 0.006  \\
10.23 & 0.98  \\
10.30 & 0.011 
\end{tabular}
\end{ruledtabular}
\end{table}

The above-mentioned estimate concerns undetected strength resulting from the insensitivity of the detector to $\gamma$-rays. Any strength associated with radiative decay would only increase the difference between the ($e$,$e^{\prime}$) result and the present. A B(M1) for the absorption can be calculated from the Moszkowski estimate to be 2.1 $\mu_N^2$ using the relationship
\begin{equation}
B(M1) = \frac{9}{16 \pi}
				\frac{2 I_x+1}{2 I_0 + 1}
				\frac{(2 m_p c^2)^2}{\alpha}
				\frac{1}{E_{\gamma}^3}
				\Gamma_{\gamma} (M1)
\end{equation}
derived from Ref.~\cite{Bohr-MottelsonVol1}. B(M1) is the transition strength for photoabsorption, $I_0 = 7/2$, and $I_x=5/2$. Though this is a substantial value, it should be considered an upper limit since the Moszkowski estimate itself is crude and likely an overestimation of the partial width because it neglects the fact that the $0f_{7/2}$ level to which the nucleus decays is nearly full. Empirically M1 photon decay is more than an order of magnitude slower than the Moszkowski estimate. Even with only an order of magnitude retardation, the photon contribution would be substantially less than our uncertainty.

The result from this work is that the B(M1) in the region of 10.23 MeV is $6.8 \pm 0.5 \mu_N^2$, a value roughly 70\% more than the previous result of Steffen \textit{et al}. \cite{Steffen83}. The quoted uncertainty is dominated by the uncertainty in the flux and the neutron detection efficiency. This difference is outside of the statistical uncertainties. The results from the ($\pi ,\pi ^{\prime }$) reaction provide some weak support for the greater suppression, see Ref. \cite{Richter85}.  

Turning to theory, the ERPA calculations of Brand \textit{et al.} \cite{Brand88,Brand90t,Brand90}, which include the 2p2h correlations, predict the localized B(M1) at 10.23 MeV to be 6.6 $\mu _{N}^{2}$ with very little additional strength in the 7.7-12.7 MeV energy range. While these calculations agree with the present results, it is expected that the inclusion of short-range correlations and a stronger coupling of 1p1h to 2p2h states would cause additional fragmentation. This could lead to enough quenching of the low-lying strength to come into agreement with the ($e$,$e^{\prime}$) results \cite{Dickhoff10}.

As mentioned in the introduction, CM at the 0$\hbar \omega$ level reduces the expected B(M1) by 25 \%. Thus for example, the total B(M1) = 8.96 $\mu_N^2$ is calculated using the code \textsc{ANTOINE} \cite{Cosel98}, which used the effective KB3 interaction \cite{KB3} and free $g$-factors $g^{free}$. This value is still 70 \% greater than the experimentally measured value and the difference is often accounted for by the rescaling of the free spin $g$-factor. \textsc{ANTOINE} predicts the total B(M1) = 5.1 $\mu_N^2$ when using the KB3 interaction but with an effective $g$-factor, $g_s^{eff} = 0.75 g_s^{free}$ \cite{Cosel98}. The result is in rough agreement with the ($e,e^{\prime}$) data.

Another SM calculation \cite{GXPF104} using the GXPF1 interaction does not predict ground state transition strengths but rather magnetic dipole moments. It reliably reproduces the experimental data for the $pf$-shell nuclei, up to Ni and Zn
 with some exceptions not including $^{48}$Ca, using the free spin and orbital $g$-factors. The ability of this $pf$-shell calculation to reproduce the experimentally measured magnetic dipole moments without scaling the free $g$-factors is similar to the calculations of the $sd$-shell.

An altogether different approach based on the TFFS extends the RPA by coupling 1p1h states to the most collectivized phonons (2p2h), and the continuum \cite{Kamerdzhiev1989}. 
These calculations are also dependent on effective spin $g$-factors. The authors found that for the $^{48}$Ca M1 resonance,  B(M1) = 8.64 $\mu_N^2$, 6.55 $\mu_N^2$, and 6.12 $\mu_N^2$ when considering coupling of 1p1h to the continuum, to 2p2h and continuum with RPA-like ground-state correlations, and to the same with additional ground-state correlations, respectively \cite{Kamerdzhiev1993,Kamerdzhiev2004}. The present result is in agreement with the calculations that include RPA-like correlations as well as coupling to the continuum and 2p2h. 

The value of the $g_s^{eff}$ hinges on whether there is a persistence of the systematic difference between the M1(spin) and the GT operators, a difference known to exist in the $sd$-shell nuclei and is associated with meson exchange currents \cite{Richter00}. It follows that if $g_{s}^{eff} \approx g_{s}^{free}$, the difference is assumed to persist in the $pf$-shell and if $g_{s}^{eff} \approx 0.75 g_{s}^{free}$, it vanishes. The work of Towner provides some support for the latter conclusion \cite{Towner87}. 

Finally, the Monte-Carlo SM calculations \cite{Koonin98} quench the spin operator (for both the GT and spin part of the M1) by 0.77, intermediate to the cases above, but far closer to the lower value as it appears as a square in the B(M1). The variance in the theoretical results is an indication of the uncertainty of the magnitude of ``beyond 0$\hbar \omega $ effects" and provides further incentive to understand the difference between the present results and those from ($e$,$e^{\prime}$).

Turning to the remainder of the ($\gamma$,$n$) excitation function, these data provide the highest resolution study using real photons above the particle emission threshold in existence and are in near agreement with the data of Ref.~\cite{OKeefe87}, see Fig. \ref{xsplot}. A comparison of $dB(E1)/dE$ data \cite{Strauch99} converted to total photoabsorption cross section, in accordance with Eq. (9) of Ref.~\cite{Gordon77}, has also been included under the assumption of purely E1 transitions. The conversion was computed for each energy bin of the $dB(E1)/dE$ data using a $B(E1)$ value associated with the energy of the bin center, obtained by integration of $dB(E1)/dE$ over the 200 keV bin width. Excluding the region between $E_{\gamma}$ = 12 and 13 MeV, the results are in accord with the present data. 
The experimental structure can be compared to that predicted by microscopic calculations. 
Both the calculations of Brand \textit{et al.} (see Fig. 15d of Ref.~\cite{Brand90}) and Kamerdzhiev \textit{et al.} (see Fig. 3.2 of Ref.~ \cite{Kamerdzhiev2004}) show structure commensurate to what is observed.
The experimentally observed plateaus at 12.5 and 14.5 MeV have corresponding structures in the ERPA calculations. More fracturing of the strength leading to a substantially smoother response does not seem to be indicated.

\section{Conclusions}
Nearly monoenergetic $\gamma$-ray beams have been utilized to study the M1 and E1 strengths in $^{48}$Ca between 9.5 and 15.3 MeV. The examined energy region includes the dominant M1 fragment at 10.23 MeV and the leading edge of the GDR. The B(M1) was measured to be $6.8 \pm 0.5 \mu_{N}^{2}$, a value substantially greater than that measured with ($e$,$e^{\prime}$). The result has multiple implications, the first being that the quenching in $^{48}$Ca is similar to that found in ERPA calculations that do not include the effects of short-range correlations. The second is that the difference could be due to an effective spin operator that is intermediate to those used in the GXPF1 interaction and KB3 interactions. Finally, it implies that meson exchange effects generating a difference in the GT and spin M1 are still somewhat active in $^{48}$Ca. Since at present this difference is not understood, further experimental work is required.


\section{Acknowledgments}

We would like to thank the National Superconducting Cyclotron Laboratory for
loaning the $^{48}$Ca target. We would also like to acknowledge discussions and
communications with Drs. B.A. Brown, I. Towner, W. H. Dickhoff, A.
Tamii, and S. Strauch. This work was supported by U.S. Department of Energy, Division of
Nuclear Physics, under grants Nos. DE-FG52-06NA26155, DE-FG02-97ER41033,
DE-FG02- 87ER-40316, and NSF/DHS Grant No. 2008-DN-077-ARI014.

\bibliography{biblio}

\end{document}